\documentclass[aps,prl,twocolumn,showpacs,superscriptaddress]{revtex4-1}

\usepackage{graphicx}
\usepackage{amsmath}
\usepackage{bm}
\usepackage{longtable}
\usepackage{xspace}

\newcommand{\cmo}{CaMn$_7$O$_{12}$}

\begin{document}

\title{Modulated spin helicity stabilized by incommensurate orbital density waves in a quadruple perovskite manganite.}

\author{R. D. Johnson}
\email{roger.johnson@physics.ox.ac.uk}
\affiliation{Clarendon Laboratory, Department of Physics, University of Oxford, Oxford, OX1 3PU, United Kingdom}
\affiliation{ISIS facility, Rutherford Appleton Laboratory-STFC, Chilton, Didcot, OX11 0QX, United Kingdom}
\author{D. D. Khalyavin}
\author{P. Manuel}
\affiliation{ISIS facility, Rutherford Appleton Laboratory-STFC, Chilton, Didcot, OX11 0QX, United Kingdom}
\author{A. Bombardi}
\affiliation{Diamond Light Source, Harwell Science and Innovation Campus, Didcot, OX11 0DE, United Kingdom}
\author{C. Martin}
\affiliation{Laboratoire CRISMAT, ENSICAEN, UMR F-6508 CNRS, 6 Boulevard du Marechal Juin, F-14050 Caen Cedex, France}
\author{L. C. Chapon}
\affiliation{Institut Laue-Langevin, BP 156X, 38042 Grenoble, France}
\author{P. G. Radaelli}
\affiliation{Clarendon Laboratory, Department of Physics, University of Oxford, Oxford, OX1 3PU, United Kingdom}

\date{\today}

\begin{abstract}
Through a combination of neutron diffraction and Landau theory we describe the spin ordering in the ground state of the quadruple perovskite 
manganite CaMn$_7$O$_{12}$ --- a magnetic multiferroic supporting an 
incommensurate orbital density wave that onsets above the magnetic 
ordering temperature, $T_\mathrm{N1}$ = 90~K. The multi-$\mathbf{k}$ magnetic structure in the ground state was found to be a nearly-constant-moment helix with modulated spin 
helicity, which oscillates in phase with the orbital occupancies on the Mn$^{3+}$ sites via trilinear magneto-orbital coupling.  Our phenomenological model also shows that, above $T_\mathrm{N2}$ = 48~K, the primary magnetic order parameter is locked into the orbital wave by an admixture of helical and collinear spin density wave structures. Furthermore, our model naturally explains the lack of a sharp dielectric anomaly at $T_\mathrm{N1}$ and the unusual temperature dependence of the electrical polarisation.

\end{abstract}


\maketitle

The interplay between charge, magnetism and the crystal lattice in transition metal oxides is one of the foundational topics in condensed matter physics, and has defined entire fields of research, including, most notably, high-temperature superconductivity.  In manganites with the simple perovskite structure (general formula $A_{1-x}A'_x$MnO$_3$, where $A$ = La, rare earth or another tri-valent metal and $A'$ = alkaline earth), this interplay results in electronic ordering of a peculiar and distinctive form, which has been understood since the 1950's thanks to the seminal neutron scattering work by Wollan and Koehler \cite{wollan55} and its interpretation by Goodenough \cite{goodenough55}. Insulating manganites typically undergo structural distortions indicative of orbital ordering (OO) and often Mn$^{3+}$/Mn$^{4+}$ charge ordering (CO) and, on further cooling, order antiferromagnetically (AFM) via super exchange. The resulting magnetic structures can be complex, but they invariably reflect and lock into the pattern of Heisenberg exchange interactions defined by the OO and CO. The quadruple perovskite manganites (general formula $A$Mn$_7$O$_{12}$, where $A$ = alkali or alkaline earth metal) display much of the same CO and OO physics as the prototypical simple perovskite manganite. For example, NaMn$_7$O$_{12}$ and La$_{0.5}$Ca$_{0.5}$MnO$_3$ both contain 50\% Mn$^{3+}$ and 50\% Mn$^{4+}$, and have very similar OO and AFM structures of the so-called C-E type \cite{radaelli97,prodi04,streltsov14,prodi14}. Our recent discovery of a very different mode of interplay in the closely related compound CaMn$_7$O$_{12}$ \cite{perks2012} demonstrates that the quadruple perovskites can play host to even more exotic electronic textures.

On cooling from 400~K, the B-site manganese ions in CaMn$_7$O$_{12}$ undergo gradual CO of nominally 75\% Mn$^{3+}$ and 25\% Mn$^{4+}$ on the $9d$ and $3b$ Wyckoff positions of space group $R\bar{3}$, respectively, and become fully ordered below room temperature \cite{przenioslo02}. The A-sites have an ordered occupation of $\tfrac{1}{4}$ Ca$^{2+}$ and $\tfrac{3}{4}$ Mn$^{3+}$ at all temperatures. Below T$_\mathrm{OO}$=250~K, an incommensurate structural modulation appears (propagation vector \textbf{k}$_\mathrm{s} = (0, 0, 2.075)$ at 150~K), associated with a continuous variation in Mn-O bond lengths \cite{slawinski2009}.  This arises from a unique incommensurate orbital density wave, in which the occupation of the B-site Mn$^{3+}$ $d_{x^2-r^2}$ and $d_{y^2-r^2}$ orbitals vary sinusoidally along the $c$ direction \cite{perks2012}. Below ${T_\mathrm{N1}}$ = 90~K, the manganese magnetic moments order in an incommensurate helical pattern with propagation vector \textbf{k}$_0 = (0, 0, 1.039)$, which breaks inversion symmetry allowing coupling to the crystal structure via the relativistic antisymmetric exchange to yield an electrical polarisation \cite{johnson2012}. As in the simple perovskite manganites, the magnetic periodicity is locked into exactly twice the lattice periodicity (\emph{i.e.} \textbf{k}$_\mathrm{s}$ = 2\textbf{k}$_0$) for $T_\mathrm{N2}<T<T_\mathrm{N1}$, where $T_\mathrm{N2}$ = 48~K \cite{slawinski10}. However, this similarity is found to be superficial by the facts that simple magneto-elastic coupling vanishes for a proper helix magnetic structure, and that the structural and magnetic propagation vectors become incommensurate with each other below $T_\mathrm{N2}$ \cite{slawinski2012}. This leaves a simple but important question unanswered:  how are spins and orbitals coupled in CaMn$_7$O$_{12}$?

In this Letter, we provide an answer based on the results of synchrotron and neutron diffraction data on powders and single crystals. Firstly, we provide direct experimental evidence for a magnetic order component with $\mathbf{k}=3\mathbf{k}_0$ in the lock-in phase between $T_\mathrm{N2}<T<T_\mathrm{N1}$, which has been predicted on pure symmetry grounds \cite{perks2012} and based on first-principle calculations \cite{cao15}. Secondly, we demonstrate that in the ground state magnetic structure ($T<{T_\mathrm{N2}}$), previously described as a ``beating'' of the high-temperature phase \cite{slawinski2012}, the fundamental magnetic order parameter becomes de-locked from the orbital modulation, which remains coupled to the magnetism through multiple additional orders. This unique form of coupling involves an infinite set of propagation vectors, four of which were directly observed in our neutron data. Together, they describe a constant-moment magnetic structure with modulated spin helicity \cite{helicity}. Finally, Landau theory is employed to fully explain the phenomenological coupling between spins and orbitals observed in all ordered phases of \cmo, and its connection with the unusual dielectric properties of this material.

\begin{figure}
\includegraphics[width=0.48\textwidth]{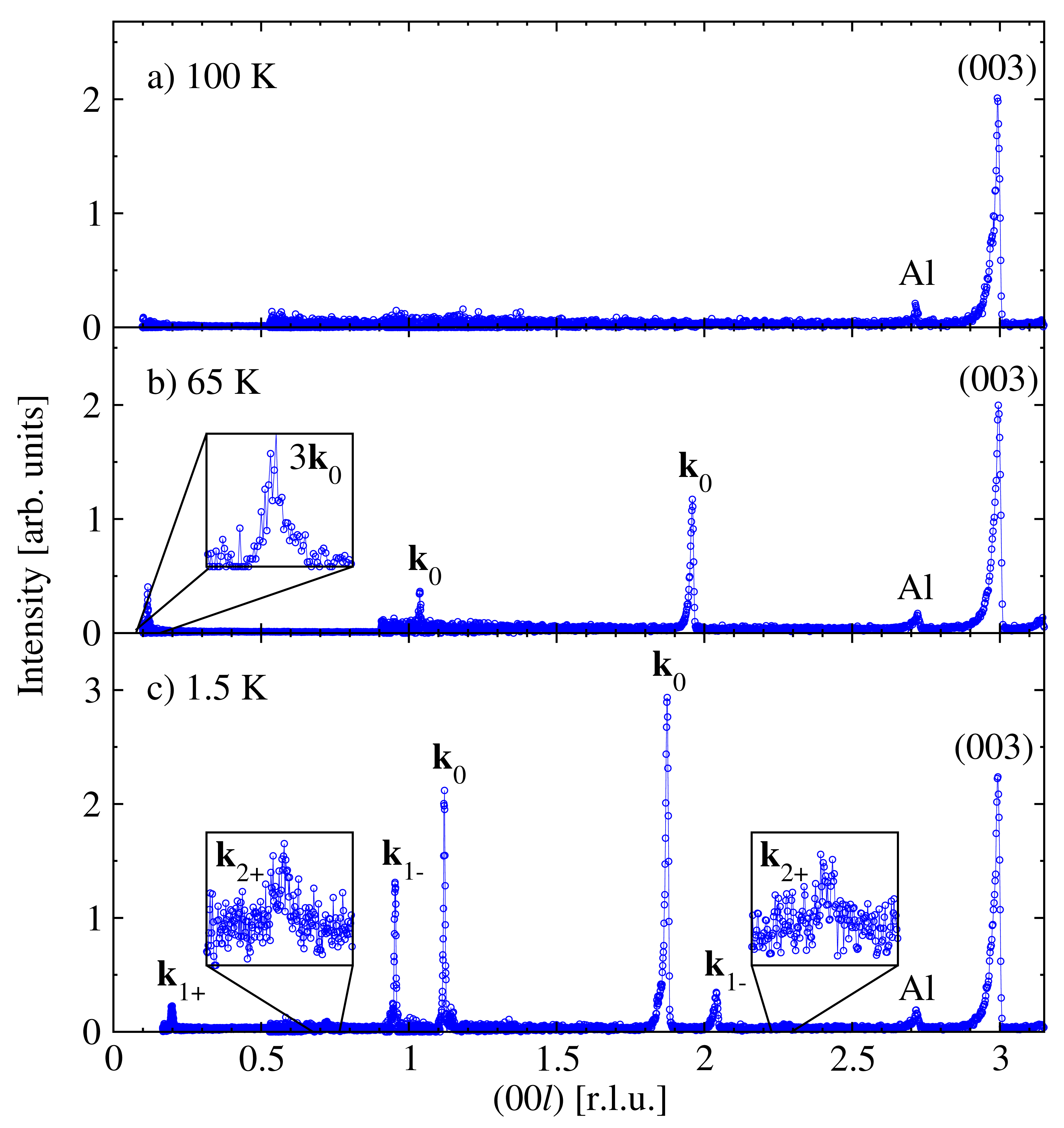}
\caption{\label{fig:SX}(Color online) Single crystal time-of-flight (TOF) neutron diffraction data measured along the $(00l)$ reciprocal space axis in each magnetic phase. At each temperature data were collected at three sample orientations corresponding to $(00l)$ two-theta angles of 12$^\circ$, 40$^\circ$, and 80$^\circ$, giving access to $d$-spacing ranges of approximately 5-40~$\mathrm{\AA}$, 2-12~$\mathrm{\AA}$, and 1-5~$\mathrm{\AA}$, respectively. Note that the TOF neutron flux profile leads to the step changes in the background once data are combined.}
\end{figure}

Single crystal neutron diffraction measurements were performed on WISH \cite{chapon11}, a time-of-flight diffractometer at ISIS, UK. An approx. $0.01\mathrm{mm}^3$ un-twinned sample was grown by the flux method (the same sample studied in references \citenum{perks2012} and \citenum{johnson2012}), and was mounted within a $^4$He cryostat and aligned with the hexagonal $c$-axis in the horizontal scattering plane. Data measured along the $(00l)$ reciprocal space zone axis are shown in Figure \ref{fig:SX} for all three magnetic phases. The $(003)$ Bragg reflection from the crystal structure and a small aluminium peak at $d=2.34\mathrm{\AA}$ originating from the sample mount were present at all temperatures. In the paramagnetic phase no additional reflections were found. At 65 K two magnetic reflections at $l=\mathbf{k}_0$ and $l=3-\mathbf{k}_0$ were observed as expected for the reported magnetic structure \cite{johnson2012}. In addition, a magnetic peak at $l=-3+3\mathbf{k}_0$ ($d=54.6 \mathrm{\AA}$) was found, determining that the helical magnetic structure between $T_\mathrm{N1}$ and $T_\mathrm{N2}$ is in fact more complex than proposed before \cite{johnson2012}. In the ground state, no less than seven magnetic diffraction peaks were observed in the range $0<l<3$, which correspond to four distinct propagation vectors as listed in Table \ref{ktab} (note that \textbf{k}$_{0}$ and \textbf{k}$_{1-}$ are equivalent to those previously reported \cite{johnson2012,slawinski2012}).  We include a 5$^{th}$ vector, \textbf{k}$_{2-}$, which was not observable within the limits of the diffractometer, but is predicted in the theoretical discussion that follows. 

\begin{table}
\caption{\label{ktab}Structural and magnetic propagation vectors of the two magnetically ordered phases of \cmo . Values determined by fitting powder diffraction data are given with standard errors.}
\begin{ruledtabular}
\begin{tabular}{c|ccc}
Temperature & \multicolumn{3}{c}{Leading-order propagation vectors}\\
\hline
& Label & Components & Fundamental $k_z$ \\
65 K & \bf{k}$_\mathrm{s}$ & - & 2.0788(2)\\
& \bf{k}$_0$ & $\tfrac{1}{2}$\bf{k}$_\mathrm{s}$ & 1.03942(9) \\
& 3\bf{k}$_0$ & 3\bf{k}$_0$ & 0.1162(2) \\
\hline
2 K & \bf{k}$_\mathrm{s}$ & - & 2.0775(1) \\
 & \bf{k}$_0$ & - & 1.12354(8) \\
 & \bf{k}$_{1+}$ & \bf{k}$_\mathrm{s} +$\bf{k}$_0$& 0.2031(4) \\
 & \bf{k}$_{1-}$ &\bf{k}$_\mathrm{s}- $\bf{k}$_0$ & 0.9554(4) \\
 & \bf{k}$_{2+}$ & 2\bf{k}$_\mathrm{s}+$\bf{k}$_0$ & 0.7215 \\
 & \bf{k}$_{2-}$ & 2\bf{k}$_\mathrm{s}-$\bf{k}$_0$ & 0.0315 
\end{tabular}
\end{ruledtabular}
\end{table}

\begin{figure}
\includegraphics[width=0.48\textwidth]{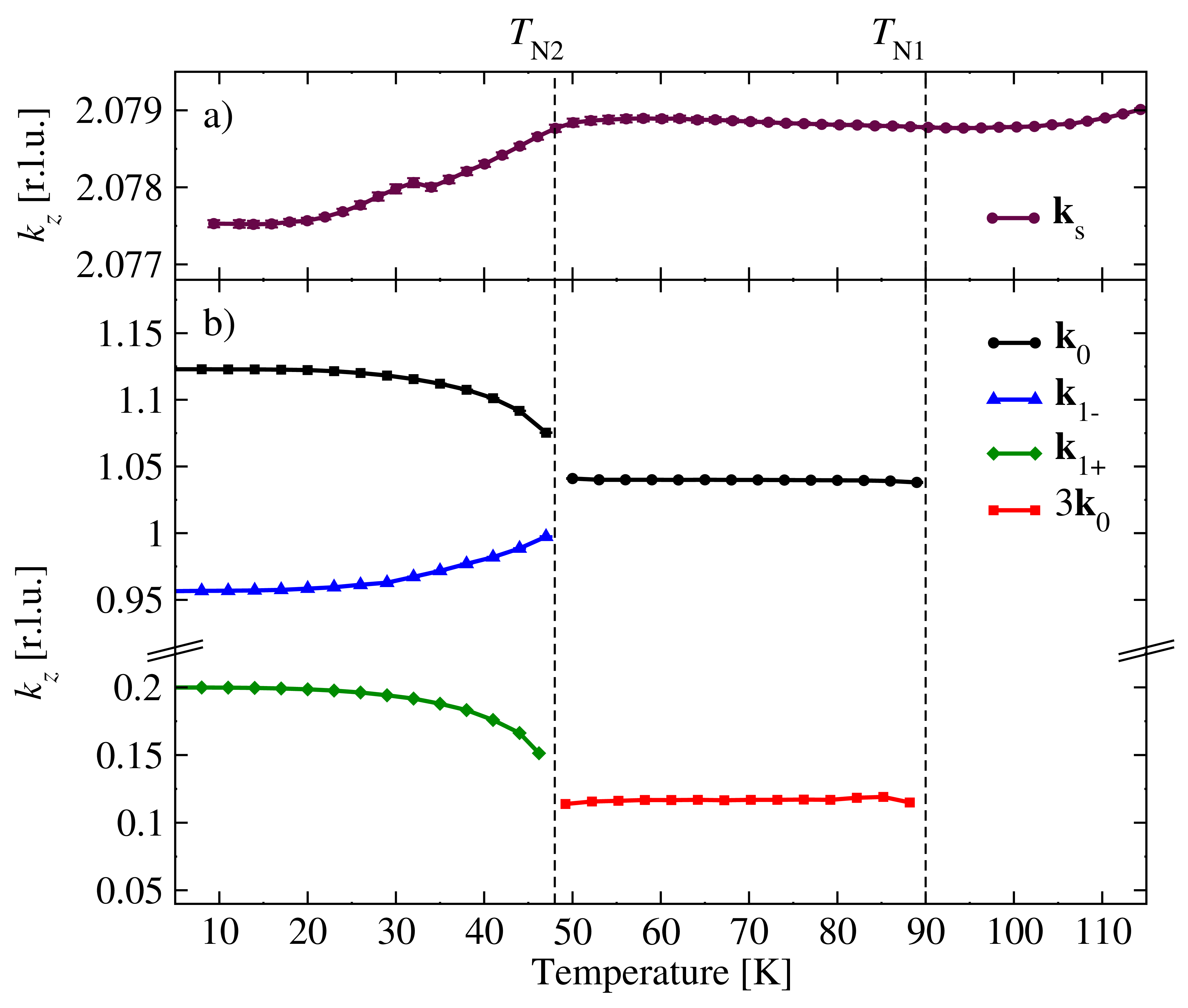}
\caption{\label{fig:temp_dep}(Color online) The temperature dependence of the $z$ components of a) the structural modulation propagation vector, and b) the magnetic modulation propagation vectors observable in neutron powder diffraction data; the \textbf{k}$_{2+}$ and \textbf{k}$_{2-}$ vectors could not be reliably measured from the powder data as their respective peak intensities were found to be too weak.}
\end{figure}

The temperature dependence of the $z$-component of \textbf{k}$_\mathrm{s}$ is plotted in Figure \ref{fig:temp_dep}a, determined through single crystal synchrotron X-ray diffraction measurements performed at I16, Diamond Light Source, UK. Similarly, Figure \ref{fig:temp_dep}b shows the $z$-component of the \textbf{k}$_{0}$, \textbf{k}$_{1+}$, \textbf{k}$_{1-}$, and 3\textbf{k}$_{0}$ vectors evaluated as a function of temperature by a systematic LeBail fit to neutron powder diffraction data measured using WISH (a 1~g powder sample was prepared by grinding single crystals and sieving through a 35 $\mu$m mesh). Below $T_\mathrm{N1}$, the fundamental magnetic modulation, \textbf{k}$_0$, locks into the structural modulation (2\textbf{k}$_0$~=~\textbf{k}$_\mathrm{s}$), and both remain approximately constant on cooling down to $T_\mathrm{N2}$. Below $T_\mathrm{N2}$ the fundamental magnetic order persists, but decouples from \textbf{k}$_\mathrm{s}$ allowing its propagation vector to increase monotonically such that at 2 K manganese spins separated by one unit cell along along the $c$-axis are rotated by an average $15^\circ$ helical rotation per atom, compared to $5^\circ$ at 65 K. Also below $T_\mathrm{N2}$, \textbf{k}$_{1+}$ and \textbf{k}$_{1-}$ onset with the strict relationships $\mathbf{k}_\mathrm{1+}=\mathbf{k}_\mathrm{s}+\mathbf{k}_0$, and $\mathbf{k}_\mathrm{1-}=\mathbf{k}_\mathrm{s}-\mathbf{k}_0$.

\begin{figure}
\includegraphics[width=0.48\textwidth]{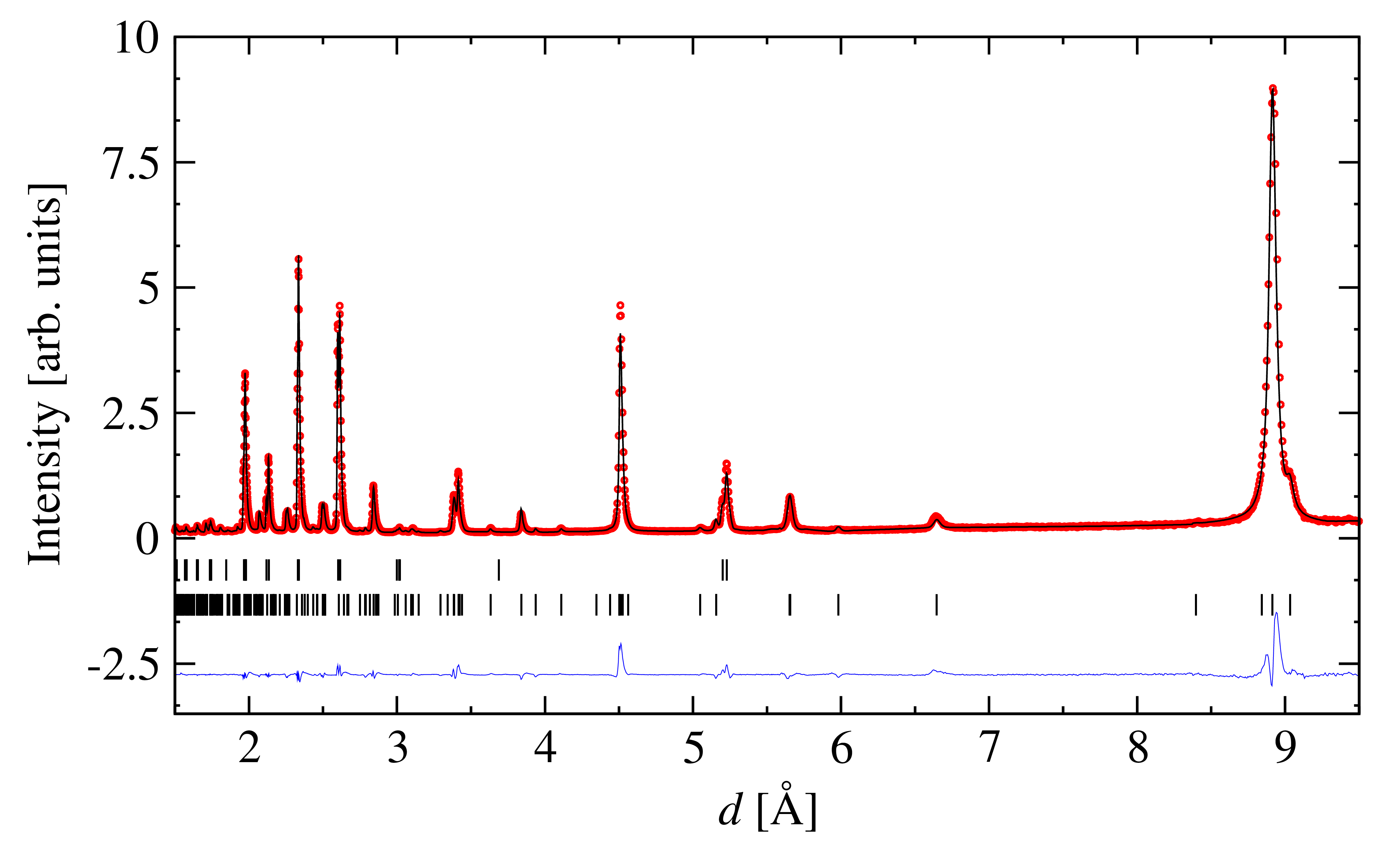}
\caption{\label{npd} (Color online) Fit to neutron powder diffraction data measured at 1.5~K. Data are shown as red circles, the fit ground state magnetic structure model as a solid black line, top and bottom tick marks indicate the position of nuclear and magnetic peaks, respectively, and the blue curve gives the difference pattern, $I_\mathrm{obs}-I_\mathrm{calc}$.}
\end{figure}

The magnetic structure between $T_\mathrm{N1}$ and $T_\mathrm{N2}$ was previously solved assuming a single propagation vector \citep{johnson2012}, as described in detail in the Supplemental Material \cite{supmat}. The presence of an additional 3\textbf{k}$_{0}$ order demonstrates greater complexity, but could not be completely solved since only a single diffraction peak was observed. This structure, however, could be inferred from that of the ground state (see below). In the ground state, Rietveld refinement and simulated annealing \cite{rodriguezcarvaja93} against high resolution neutron powder diffraction data (Figure \ref{npd}) showed that all five magnetic propagation vectors correspond to modulations with the same basic structure as the fundamental order, \textbf{k}$_0$ \cite{johnson2012}, \emph{i.e.} all magnetic orders independently take the form of a simple helical structure, differing only in helicity and magnitude.  Due to limitations inherent to powder data, we refined a model with relative phases and amplitudes constrained in such a way as to yield an \emph{exactly constant-moment structure}, described explicitly in the Supplemental Material \cite{supmat}. The moment magnitudes on all three manganese sublattices associated with the five propagation vectors are given in Table \ref{momtab}, which reveals that the largest modulation of spin helicity at 1.5 K occurs on the Mn$^{3+}$ B-sites. This model provides an excellent fit to the data in Figure \ref{npd}, up to a small discrepancy in the relative intensities of the $\mathbf{k}_\mathrm{1+}$ and $\mathbf{k}_\mathrm{1-}$ magnetic peaks. This additional subtlety can be understood in the context of the Landau theory discussed below.

\begin{table}
\caption{\label{momtab}Magnetic moment magnitudes on each manganese sublattice. \textbf{k}$_0$ and (\textbf{k}$_{1+}$ \& \textbf{k}$_{1-}$) were refined against neutron powder diffraction data. Moment magnitudes for (\textbf{k}$_{2+}$ \& \textbf{k}$_{2-}$) were calculated as the diffraction signals were too weak for reliable fitting. N.B. Predicted values were found to be fully compatible with the experimental data.}
\begin{ruledtabular}
\begin{tabular}{c|ccc}
&\multicolumn{3}{c}{Moment ($\mu_\mathrm{B}$)} \\
\cline{2-4}
Modulation & A-site Mn$^{3+}$ & B-site Mn$^{3+}$ & B-site Mn$^{4+}$ \\
\hline
\bf{k}$_0$ & 3.71(3) & 2.69(2) & 3.09(6)\\
\bf{k}$_{1+}$ \& \bf{k}$_{1-}$ & 0.39(2) & 1.41(2) & 0.52(4)\\
\bf{k}$_{2+}$ \& \bf{k}$_{2-}$ & 0.02 & 0.35 & 0.04\\
\hline
\bf{Total}: & 3.7(2) & 3.1(1) & 3.2(3)
\end{tabular}
\end{ruledtabular}
\end{table}

\begin{figure}
\includegraphics[width=0.48\textwidth]{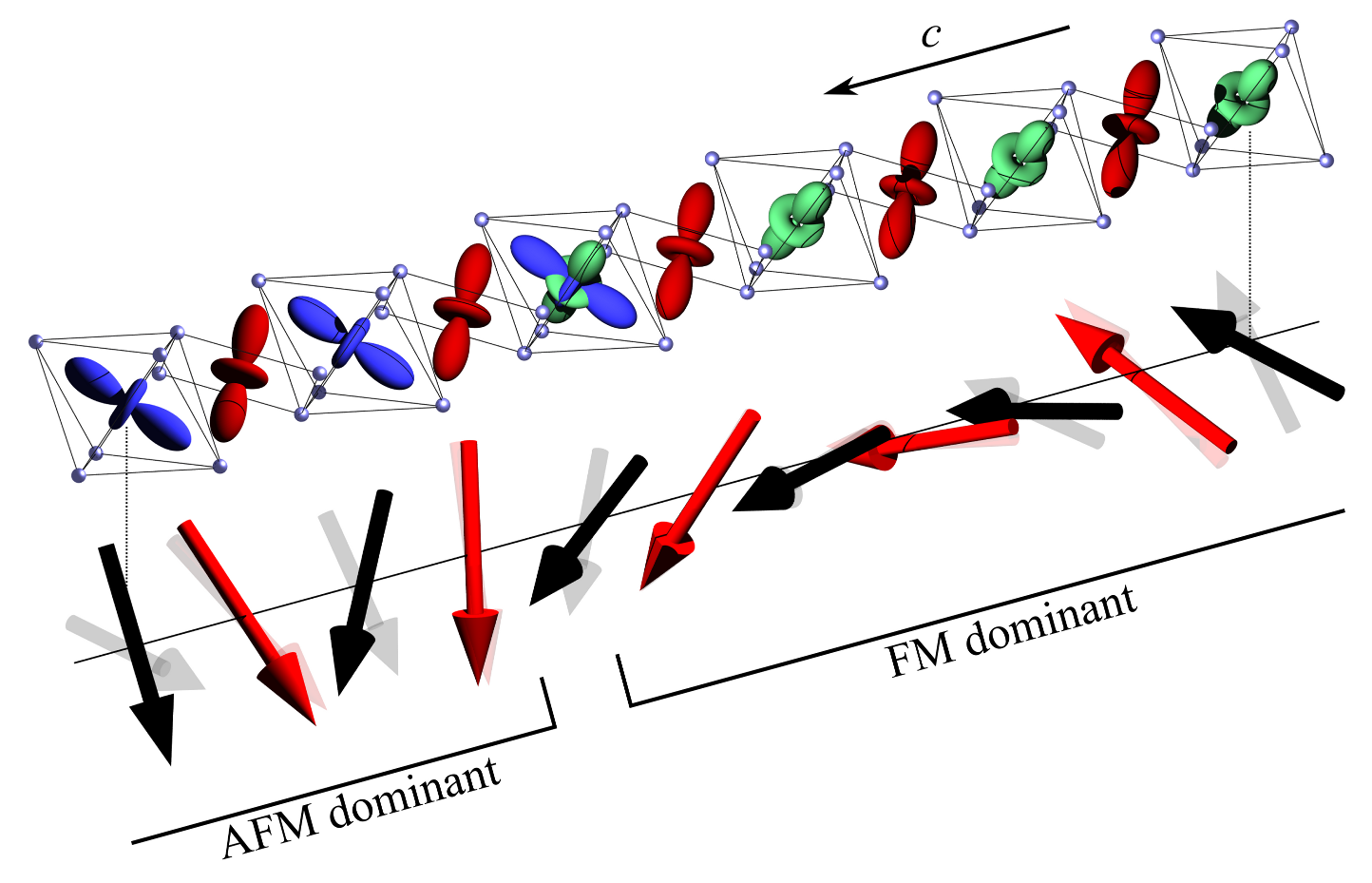}
\caption{\label{magstrucfig}(Color online) The ground state magnetic structure of \cmo\ projected along the $c$-axis.The spins at the manganese sites at (0.5,0,$z$), $z=n$, and $n+\tfrac{1}{2}$ are shown parallel to a cartoon of the orbital density waves projected onto the same sites. A-site Mn$^{3+}$, B-site Mn$^{3+}$, and B-site Mn$^{4+}$ spins are shown in red, black, and yellow, respectively. B-site Mn$^{3+}$ $d_{x^2-r^2}$ and $d_{y^2-r^2}$ orbitals are shown in blue and green, and the magnetic structure without helical modulation is superimposed in transparency. Note that the exact phase between orbital and magnetic order is not determined empirically.}
\end{figure}

Figure \ref{magstrucfig} shows the magnetic moments of A-site and B-site Mn$^{3+}$ ions at (0.5,0,$z$), where $z=n$ and $n+\tfrac{1}{2}$, respectively, drawn extended along the direction of propagation (hexagonal $c$-axis), and parallel to a cartoon of the orbital density waves projected onto the same (0.5,0,$z$) sites. The helicity of the constant-moment magnetic structure is modulated by the same periodicity of the orbital density wave, $\mathbf{k}_\mathrm{s}$, and can be qualitatively understood by employing the well-known Goodenough-Kanamori-Anderson rules \cite{anderson50,kanamori57,goodenough58}. The nearest neighbour exchange interactions are mediated by the Mn$^{3+}$ $d$ orbitals \cite{orbitals}, and are therefore modulated with the same periodicity as the orbital density waves. It follows that on moving along the $c$-axis, regions of spins will alternately experience more \emph{ferromagnetic} and then more \emph{antiferromagnetic} interactions, leading to a periodic bunching of the spins and a reduced helicity, followed by a greater rotation between adjacent spins and increased helicity, respectively.

The experimentally observed high-temperature lock-in phase and transition to the multi-${\bf k}$ ground state can be fully understood based on symmetry arguments, in which $R\bar{3}1'$ is the parent space group symmetry and the incommensurate structural modulation retains the same point group symmetry $\bar{3}1'$, transforming according to the two-dimensional time-even representation (irrep) $\Lambda_1$ (Miller and Love notation, used throughout).  The associated order parameters $(\delta ,\delta^*)$ develop just below room temperature and can be considered fixed through the magnetic phase transitions. The magnetic orders are individually associated with the time-odd four-dimensional physically irreducible representation  $\Lambda_2 \oplus \Lambda_3$, with generic order parameters $(\eta,\xi^*,\xi,\eta^*)$. Matrix operators for the generating symmetry elements of $R\bar{3}1'$,  for   $\Lambda_1$ and $\Lambda_2 \oplus \Lambda_3$, respectively, are summarized in Tables I and II of the Supplemental Material \cite{supmat}.  In general, there are three symmetry-distinct order parameter directions associated with the $\Lambda_2 \oplus \Lambda_3$ irrep: 1) $(\eta,0,0,\eta^*)$ or $(0,\xi^*,\xi,0)$ describe a proper helix with opposite helicity, respectively, both with \emph{polar} $31'$ magnetic point symmetry, and similar to the one proposed for the high-temperature magnetic structure in our previous work \cite{johnson2012}.  2) $(\eta,\xi^*,\xi,\eta^*)$ with $|\eta|=|\xi|$ describes a collinear SDW with \emph{non-polar} $\bar{1}1'$ point symmetry, and 3) a combination of these two represents an admixture of proper helix and SDW, with \emph{polar} $11'$ magnetic point symmetry.

We will first consider the free-energy invariants in the lock-in phase. The lowest-order invariants that couple the magnetic order parameters to the structural modulation, $\delta$, have the form $\delta^{*} \eta_0 \xi_{0}+c.c.$ and $\delta \eta_0 \eta^*_{3}+\delta \xi_0 \xi^*_{3}+c.c$, where $c.c.$ indicates the complex conjugation, and $\eta_{0}, \xi_{0}$ and $\eta_{3}, \xi_{3}$ are components of the fundamental ($\mathbf{k}_0$) and third harmonic ($3\mathbf{k}_0$) order parameters, respectively.  Higher-order terms in $\delta^n$ are also allowed, and are described below for the general (de-locked) case. Immediately below $T_{N1}$, the term $\delta^{*} \eta_0 \xi_{0}+c.c.$ leads to spins ordering in a collinear SDW ($|\eta_0| \simeq |\xi_0|$), which is favoured by entropy, and is locked into the structural modulation with $2\mathbf{k}_0=\mathbf{k}_\mathrm{s}$. The lock-in is supported by an invariant of the form $\epsilon^* \eta_0 \xi_0^*$, where $\epsilon$ is a spontaneous triclinic strain. This invariant is forbidden by symmetry when $\mathbf{k}_\mathrm{s} \ne 2\mathbf{k}_0$, and is effectively of higher order in the spin parameters. Physically, this lock-in term expresses the tendency of the spins and associated magneto-elastic distortion to align along particular directions in the crystal structure, or their symmetry-equivalent ones.

On further cooling within the lock-in phase, a net helicity gradually develops with $|\eta_0| > |\xi_0|$ in the right-handed case (chosen throughout for clarity --- the left-handed case is in general constructed by interchanging $\eta$ with $\xi$). The system evolves towards a constant moment magnetic structure where the helicity is modulated by $\mathbf{k}_s$. This evolution is signalled by the development of $\eta_{3}$ (see Figure 2 in the Supplemental Material \cite{supmat}), where in the limit of constant moment, $|\eta_{3}|= |\xi_{0}|$. As a consequence, the lock-in energy decreases in relative magnitude, until at $T=T_\mathrm{N2}$ it equals the exchange energy lost by the lock-in.  At this point, a first-order transition occurs, and $\mathbf{k}_0$ de-locks from $\mathbf{k}_\mathrm{s}$. 

In the ground state, the free energy invariants have the general form $\delta^{n} \eta_0 \eta^*_{n+}+\delta^n \xi_0 \xi^*_{n+}+c.c.$ and $\delta^{*n} \eta_0 \xi_{n-}+\delta^{*n} \xi_0 \eta_{n-}+c.c.$, where $n=1 \dots \infty$, and the propagation vector of $\eta_{n\pm}$ and $\xi_{n\pm}$ is $\mathbf{k}_{n\pm}=\mathbf{k}_s\pm\mathbf{k}_0$. In CaMn$_7$O$_{12}$, these invariants give rise to a constant moment helix, in which the helicity is modulated \emph{incommensurate} with the fundamental helical pitch, but \emph{commensurate} with the structural modulation, as determined experimentally. Consistent with the phase transition at $T_\mathrm{N2}$, this structure is energetically favoured by magnetic exchange since it allows $\mathbf{k}_0$ to be optimised. Furthermore, even at the lowest temperatures a SDW component may persist, which might explain the small discrepancy in the relative intensities of the $\mathbf{k}_\mathrm{1+}$ and $\mathbf{k}_\mathrm{1-}$ magnetic peaks observed experimentally.

In summary, by a combination of neutron diffraction and symmetry analysis we have shown that magnetic ordering in CaMn$_7$O$_{12}$ is characterised by a continuous evolution from a non-chiral, non-polar SDW towards a constant-moment helix with modulated spin helicity, which is both chiral and polar. At the second magnetic transition temperature, $T_\mathrm{N2}$, the primary magnetic order parameter de-locks from the structural modulation, while magneto-orbital coupling persists due to additional magnetic order parameters. This model accounts for some of the most puzzling features of ferroelectricity in CaMn$_7$O$_{12}$.  In particular, the continuous development of polarity explains the absence of a dielectric anomaly at $T_\mathrm{N1}$ and the fact that the polarisation does not follow the fundamental magnetic order parameter.  Conversely, the second magnetic transition at $T_\mathrm{N2}$ is first-order, and is associated with a small but clear dielectric signature.

\begin{acknowledgments}
RDJ acknowledges support from a Royal Society University Research Fellowship. The work done at the University of Oxford was funded by an EPSRC grants, number EP/J003557/1, entitled ``New Concepts in Multiferroics and Magnetoelectrics" and number EP/M020517/1, entitled ``Oxford Quantum Materials Platform Grant''. We acknowledge Diamond Light Source for time on Beamline I16 under Proposal MT5693.
\end{acknowledgments}

\bibliography{CMO_GS}

\begin{thebibliography}{21}
\bibitem [{orb()}]{orbitals}%
  \BibitemOpen
  \href@noop {} {}\bibinfo {note} {The A-site and B-site manganese cations are
  located within square-planar and elongated octahedral oxygen coordinations,
  respectively. At both sites the crystal electric field has the same symmetry
  (D$_{4h}$) and the Mn orbital state is qualitatively identical.}\BibitemShut
  {Stop}%
\end{thebibliography}

\end{document}